# Comment on:

# Observation of the Wigner-Huntington transition to metallic hydrogen.

Paul Loubeyre[1], Florent Occelli[1] and Paul Dumas[2].

[1]*CEA, DAM, DIF, F-91297 Arpajon, France.*

[2]*Synchrotron SOLEIL, BP 48, F-91192 Gif-sur-Yvette, France.*

In a recently published article [1], Ranga P. Dias & Isaac F. Silvera have reported the visual evidence of metallic hydrogen concomitantly with its characterization at a pressure of 495 GPa and low temperatures. We have expressed serious doubts of such a conclusion when interviewed to comment on this publication [2,3]. In the following comment, we would like to detail the reasons, based on experimental evidences obtained by us and by other groups worldwide that sustain our skepticism. We have identified two main flaws in this paper, as discussed in details below: the pressure is largely overestimated; the origin of the sample reflectivity and the analysis of the reflectance can be seriously questioned.

Few groups around the world have devoted many efforts over the past 20 years to compress hydrogen above 300 GPa and to reliably characterize its properties. All have used Diamond Anvil Cells (DAC) to produce the required very high static pressure that is a prerequisite to the formation of metallic solid hydrogen. If each group has its own optimized DAC design and its favorite gasket material, all experiments have been performed using the same design for the diamond anvil, with a culet size between 20 μm – 30 μm in diameter and a diamond bevel to a diameter of 300 μm with an angle in between 8° - 11°. All attempts to reach the highest pressure possible have, up to now, been limited to a maximum value of about 380 GPa. Hydrogen is probably the most difficult system to compress, due to severe constrains: its diffusion in the gasket and in the diamond defects, its high chemical reactivity with surrounding materials and its intrinsic strong compressibility. However, these constrains cannot solely explain this apparent limitation in pressure increase since similar limitation has been faced for the compression of metals. Only recently, the development of double stage diamond anvils have successfully allowed to largely exceed the 400 GPa limit on metals [4], but compressing hydrogen in such a DAC configuration is still a formidable challenge . Accordingly, it is more than surprising that Dias & Silvera could report a compression of hydrogen up to 495 GPa, with regards to the largely shared experience on the standard bevel geometry for diamond anvils used in DAC to reach multi-Mbar pressures. They argued, in their article, that this remarkable performance is due to the etching of the diamond surface and to the use of synthetic diamond. One has to recognize that such process could be very useful in reducing or even in removing completely the number of defects on the anvils tip, that sometimes cause premature breakage of the anvil, but with no means explains a better mechanical performance of the anvils. In reality, as demonstrated below, they have , instead, reached at maximum a pressure of about 340 GPa, instead of the claimed

495 GPa. Already, as such, it is however a remarkable performance when using 30 µm in diameter culet.

Hereafter are the experimental arguments sustaining that the maximum pressure reach has been largely overestimated. The Raman shift of the diamond layer at the interface with the sample is currently used to estimate the pressure value above 200 GPa on a spectroscopic laboratory bench. The calibration of Akahama et al is commonly used [5]. In Dias&Silvera paper, the Raman spectra of the diamond anvil part in contact with the hydrogen sample was only probed at the maximum pressure reached and a measured peak was attributed to the diamond phonon giving a pressure of 495 GPa. However, the correct assignment of this peak to the diamond phonon was not tested, either by scanning the pressure distribution at the diamond tip nor by varying the pressure. Nevertheless, we could figure out the compression path of the sample by using other measurements given in the Dias&Silvera paper, namely the visual observation of the sample turning dark and the values of the infra-red $H_2$ vibron frequency. First, Dias&Silvera report that solid hydrogen starts darkening above 355 GPa and is definitely black at 400 GPa. More than 15 years ago, we had observed black hydrogen at 320 GPa [6], that is 80 GPa below the reported pressure in Dias&Silvera paper! We had also estimated the pressure using the phonon shift at the diamond tip, yet using a different calibration to calculate the pressure. We can revisit our pressure estimate using the more accurate calibration of Akahama: Black hydrogen was obtained for a diamond phonon frequency at the tip of 1816 cm$^{-1}$, that corresponds to a pressure of 300 GPa, hence even at slightly lower pressure. In addition, black hydrogen at 80K was also observed at 300 GPa by Akahama et al and in that case also the observation of black hydrogen was associated to a pressure measurement using the diamond phonon pressure scale[7]. On the other hand, Zha et al reported the darkening of hydrogen at 360 GPa, however the pressure was not directly measured but estimated using the extrapolation of lower pressure measurements [8]. This is probably the reason of this higher pressure. In the same paper, Zha et al reported the evolution of the infra-red $H_2$ vibron frequency versus pressure at 80 K . Accordingly, the pressure dependent $H_2$ vibron reported in the Zha's article is questionable due to the imprecision in determining the $H_2$ sample pressure. Dias&Silvera have used these data of Zha et al to obtain the pressure in their hydrogen sample based on the $H_2$ IR vibron frequency they measured, doing so pressure estimates in between 88 GPa and 335 GPa are presented in figure S1 of their paper. Then, in figure S3, a linear pressure evolution versus load in match with the maximum 495 GPa pressure is hence plotted. However, in figure 1 below, the $H_2$ Infra-red vibron frequency versus pressure we measured in four independent experiments, using the synchrotron bright infrared source at SMIS beamline of SOLEIL, are diverging from the values given by Dias&Silvera in their supplementary materials. More details on the experimental configuration of these infra-red measurements can be obtained in the published data at 300 K [9]. The data of Dias&Silvera are systematically at higher pressure for a given vibron frequency , with a marked departure above 250 GPa and that is due to their poor estimation of pressure using Zha measurements, as discussed above.

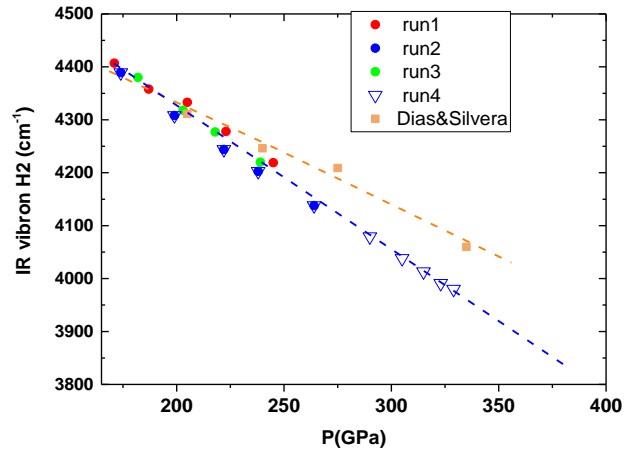

*Figure 1*: Infra-red vibron frequency of hydrogen versus pressure at 80 K. The data collected during four independent experiments carried out with a synchrotron infrared source at SOLEIL, which provide a much higher spectral quality than a thermal source especially with such small gasket holes, are compared to the data of Dias&Silvera presented in their supplementary materials. In our data, the pressure is measured using the frequency of the diamond phonon using the Akahama scale [5]

We have corrected in Figure 2 the pressure evolution estimated by Dias&Silvera, based on the real pressure of the darkening of hydrogen ( 300 GPa instead of 400 GPa) and on the pressure estimated from the IR vibron frequency pressure shift of our measurements. The evolution of pressure versus load is, accordingly, no longer linear but it has the expected sub-linear behavior, similar to what is typically obtained ( as examples, see in figure 3 below three of ours different pressurization configuration runs on hydrogen at SOLEIL) . This pressure-load behavior is explained by the change of regime from the gasket deformation to the diamond anvil deformation to make the increase of pressure, the diamond deformation at the tip being related to the flattening of the anvil bevel. As seen in figure 2, the maximum pressure reached by Dias&Silvera should be 340 GPa rather than 495 GPa. Furthermore, this maximum pressure value is in good agreement with the empirical relation proposed some years ago by Ruoff [10] to estimate the maximum pressure that can be achieved with bevel anvils having a culet of diameter D, namely: $P(GPa)= 1856\ D(\mu m)^{-1/2}$. This relation has been verified in many experiments by various groups since then. It states that with 30 microns culet with a beveled geometry, the maximum pressure should be around 340 GPa.

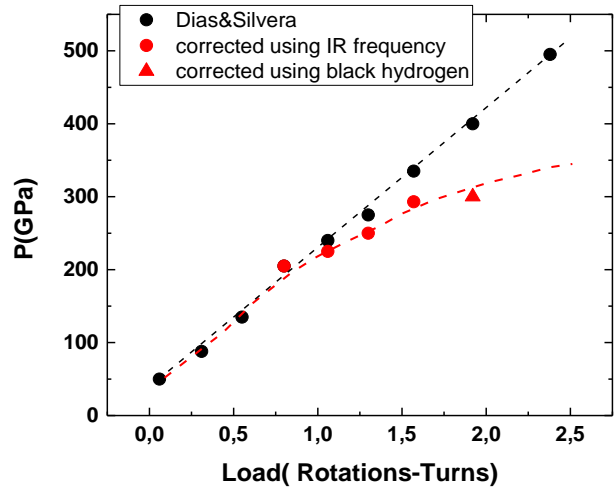

*Figure 2:* Pressure as function of load given by Dias&Silvera , as black dots, and using the corrected values of pressure, either using our pressure of observation of dark hydrogen, as the red triangle, or our measured curve of the H2 IR vibron frequency shift versus pressure, as dots. The red dash line is the interpolated pressure versus load evolution.

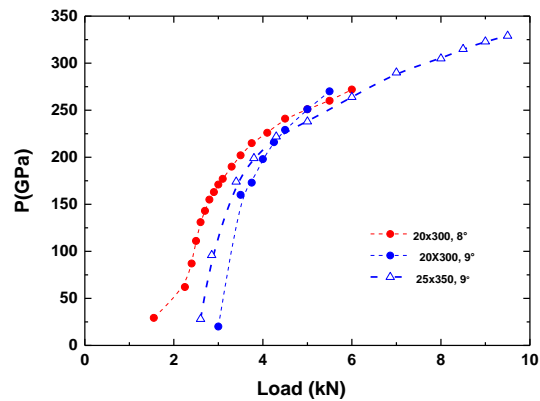

*Figure 3:* Typical pressure versus load curves obtained during our IR measurements on solid hydrogen at SOLEIL, the load in kN being proportional to the pressure in the membrane pushing the piston. Three runs with different bevel angles are shown.

The visual observation of the sample reported in the Dias&Silvera article is puzzling . At 495 GPa, the area at tip should appear very brownish. To analyze their reflectance measurements, Dias&Silvera have taken into account the attenuation properties of diamond at high pressure as measured previously by Vohra [11]. In the supplementary Materials, Dias&Silvera give in figure S4 the two-path optical density of stressed diamond used to correct the reflectance spectra. At 495 GPa, the optical density should be around 8 in the blue, 6 in the green and 4 in the red. The sample should then appear very brown dark due to the large pressure peaked at the diamond tip, and very bright for the bevel since the pressure is much lower, hence the optical properties of the diamond anvil hardly perturbed away from the culet. Thus, the visual observation of the sample, as shown in figure 2 of the Dias&Silvera paper, seems also to indicate a much lower pressure than 495 GPa.

The analysis of the reflectance based on the correction of diamond absorption at 495 GPa should also be questioned since the pressure is, in fact, about 340 GPa and therefore the absorption of diamond is totally different. But even in accepting the 495 GPa estimation of pressure made in their article, the corrected reflectance presented in figure 3a of the Dias&Silvera paper from the raw data of figure 3b using the absorption properties of figure S4 is not consistent. In order to provide numbers for this inconsistency, the analyzed reflectance of around 0.85 in between 0.8 eV and 3.0 eV should be associated to a raw reflectance data of almost zero at 3 eV rather than 0.45 eV, when taking into account the absorption of the stressed diamond with an optical density of 8 at 3 eV and 1 at 0.8 eV.

One still needs to address the question of the reflective nature of the sample. Let's suppose that hydrogen is metallic at the corrected pressure of 340 GPa, as explained above. This is at much lower pressure than expected from the most advanced calculations and the extrapolation of the closure of the experimental electronic gap[5, 13]. Furthermore, the conductivity of hydrogen at 360 GPa and 80 K has been measured by Eremets showing that hydrogen is not a metal [12]. So, to explain the visual observation reported in their article, it could well be that the reflectivity is due to the 50 nm $Al_2O_3$ layer, either chemically transformed under the action of hydrogen or because of its transition to a metallic glass in this pressure range, as suggested by Nellis [14].

The several issues pointed out here must be clarified before the claim of the achievement of metallic hydrogen can be established.